\documentclass[zpreprint,zbstdefault]{./LaTeX/zeus/zeus_paper}
\usepackage[english]{babel}

\newcommand{\ZcoosysB}{%
The ZEUS coordinate system is a right-handed Cartesian system, with the $Z$
axis pointing in the proton beam direction, referred to as the ``forward
direction'', and the $X$ axis pointing left towards the centre of HERA.
The coordinate origin is at the nominal interaction point.\xspace}
\newcommand{\Zpsrap}{%
The pseudorapidity is defined as $\eta=-\ln\left(\tan\frac{\theta}{2}\right)$,
where the polar angle, $\theta$, is measured with respect to the proton beam
direction.\xspace}

\newcommand{\ZcoosysfnBeta}{\footnote{\ZcoosysB\Zpsrap}}
\newcommand{\Zdetdesc}{%
A detailed description of the ZEUS detector can be found 
elsewhere~\cite{zeus:1993:bluebook}. A brief outline of the 
components that are most relevant for this analysis is given
below.\xspace}
\newcommand{\Zctddesc}[1]{%
Charged particles are tracked in the central tracking detector
(CTD)~\citeCTD, which operates in a magnetic field of $1.43\Tesla$
provided by a thin superconducting solenoid. The CTD consists of
72~cylindrical drift chamber layers, organised in nine superlayers
covering the polar-angle#1 region
\mbox{$15^\circ<\theta<164^\circ$} ($2.02 > \eta >-1.96$). 
The transverse-momentum resolution for full-length tracks is
$\sigma(p_T)/p_T=0.0058p_T\oplus0.0065\oplus0.0014/p_T$, with $p_T$ in
$\Gev$.}
\newcommand{\Zcaldesc}{%
The high-resolution uranium--scintillator calorimeter (CAL)~\citeCAL
consists of three parts: the forward (FCAL, $1.1<\eta<3.8$), the
barrel (BCAL, $-0.7<\eta <1.1$) and the rear (RCAL, $-3.4<\eta
<-0.7$) calorimeters. Each part is subdivided transversely into towers
and longitudinally into one electromagnetic section (EMC) and either
one (RCAL) or two (BCAL and FCAL) hadronic sections (HAC). The
smallest subdivision of the calorimeter is called a cell. In the EMC
section, the towers are divided transversely into either four (FCAL
and BCAL) or two (RCAL) cells. The CAL energy resolutions, as measured
under test-beam conditions, are $\sigma(E)/E=0.18/\sqrt{E}$ for
electrons and $\sigma(E)/E=0.35/\sqrt{E}$ for hadrons, with $E$ in
$\Gev$.}
\chardef\usc=95
\chardef\til=126
\catcode`\@=11 % @ signs are now treated as letters
\DeclareRobustCommand\xdotspace{\futurelet\@let@token\@xdotspace}
\def\@xdotspace{%
  \ifx\@let@token.\else
  \ifx\@let@token\bgroup.\else
  \ifx\@let@token\egroup.\else
  \ifx\@let@token\/.\else
  \ifx\@let@token\ .\else
  \ifx\@let@token~.\else
  \ifx\@let@token!.\else
  \ifx\@let@token,.\else
  \ifx\@let@token:.\else
  \ifx\@let@token;.\else
  \ifx\@let@token?.\else
  \ifx\@let@token/.\else
  \ifx\@let@token'.\else
  \ifx\@let@token).\else
  \ifx\@let@token-.\else
  \ifx\@let@token\@xobeysp.\else
  \ifx\@let@token\space.\else
  \ifx\@let@token\@sptoken.\else
   .\space
   \fi\fi\fi\fi\fi\fi\fi\fi\fi\fi\fi\fi\fi\fi\fi\fi\fi\fi}
\catcode`\@=12 % @ signs are no longer letters

\newcommand{\stru}[2]{%
   \relax\ifmmode\hbox{\vrule height#1 depth#2 width0pt}%
   \else\vrule height#1 depth#2 width0pt\fi}
\newcommand{\Ronum}[1]{\uppercase\expandafter{\romannumeral#1}}
\newcommand{\ronum}[1]{\expandafter{\romannumeral#1}}
\DeclareRobustCommand{\LaTeXZ}{%
  \LaTeX\kern-.05em4\kern-.1em
  {\raisebox{-0.2ex}{$\scriptstyle\text{ZEUS}$}}\xspace}
\newcommand{\fig}[1]{Fig.~\ref{fig-#1}}

\DeclareMathAlphabet{\mathbf}{OT1}{cmr}{bx}{sl}
\newcommand{\eVdist}{\kern-0.06667em}

\newcommand{\Gev}{{\text{Ge}\eVdist\text{V\/}}}

\newcommand{\gev}{{\,\text{Ge}\eVdist\text{V\/}}}

\newcommand{\cm}{\,\text{cm}}

\newcommand{\Tesla}{\,\text{T}}

\newcommand{\slashfrac}[2]{%
  \raisebox{0.5ex}{\ensuremath #1}\kern-0.12em/\kern-0.08em
  \raisebox{-.8ex}{\ensuremath #2}}
\newcommand{\sqr}[3]{%
    {\vcenter{\hrule height.#3ex\hbox{\vrule width.#2ex height#1ex
     \kern#1ex\vrule width.#3ex}\hrule height.#2ex}}}

\catcode`\@=11 % @ signs are now treated as letters
\newcommand{\parenbar}{\mathpalette\p@renb@r}
\def\p@renb@r#1#2{\vbox{%
  \ifx#1\scriptscriptstyle \dimen@.7em\dimen@ii.2em\else
  \ifx#1\scriptstyle \dimen@.8em\dimen@ii.25em\else
  \dimen@1em\dimen@ii.4em\fi\fi \offinterlineskip
  \ialign{\hfill##\hfill\cr
    \vbox{\hrule width\dimen@ii}\cr
    \noalign{\vskip-.3ex}%
    \hbox to\dimen@{$\mathchar300\hfil\mathchar301$}\cr
    \noalign{\vskip-.3ex}%
    $#1#2$\cr}}}
\catcode`\@=12 % @ signs are no longer letters

\newcommand{\IP}{{\rm I$\kern-0.01667em$P}\xspace}

\mathchardef\qsm=63
\mathchardef\pls=43
\mathchardef\mns=512
\mathchardef\plm=518
\mathchardef\eql=61
\mathchardef\smallleft=300
\mathchardef\smallright=301
\mathchardef\les=316
\mathchardef\gre=318
\mathchardef\leq=532
\mathchardef\grq=533
\catcode`\@=11 % @ signs are now treated as letters
\newcounter{pict@width}
\newcounter{pict@height}
\newlength{\pict@scale}
\newcommand{\psfigadd}[4]{%
\setcounter{pict@width}{1*\ratio{#2+\pict@scale/2}{\pict@scale}}
\setcounter{pict@height}{1*\ratio{#3+\pict@scale/2}{\pict@scale}}
\setlength{\unitlength}{\pict@scale}
\hbox to #2{\hspace{-\fill}\begin{picture}(\thepict@width,\thepict@height)
\put(0,0){\psfig{figure=#1,width=#2,height=#3,clip=}}
\SetScale{0.283466457}
\SetWidth{1.763889}
{#4}
\end{picture}}
}
\newcounter{pict@widthfst}
\newcounter{pict@widthscd}
\newcounter{pict@widthtot}
\newcommand{\psfigaddtwo}[7]{%
\setcounter{pict@widthfst}{1*\ratio{#2+\pict@scale/2}{\pict@scale}}
\setcounter{pict@widthscd}{1*\ratio{#2+#4+\pict@scale/2}{\pict@scale}}
\setcounter{pict@widthtot}{1*\ratio{#2+#4+#6+\pict@scale/2}{\pict@scale}}
\setcounter{pict@height}{1*\ratio{#3+\pict@scale/2}{\pict@scale}}
\setlength{\unitlength}{\pict@scale}
\hbox{\hspace{-\fill}\begin{picture}(\thepict@widthtot,\thepict@height)
\put(0,0){\psfig{figure=#1,width=#2,height=#3,clip=}}
\put(\thepict@widthscd,0){\psfig{figure=#5,width=#6,height=#3,clip=}}
\SetScale{0.283466457}
\SetWidth{1.763889}
{#7}
\end{picture}}
}
\newcommand{\psfigror}[4]{%
\setcounter{pict@width}{1*\ratio{#2+\pict@scale/2}{\pict@scale}}
\setcounter{pict@height}{1*\ratio{#3+\pict@scale/2}{\pict@scale}}
\setlength{\unitlength}{\pict@scale}
\hbox{\begin{picture}(\thepict@width,\thepict@height)
\put(0,\thepict@height){\psfig{figure=#1,width=#3,height=#2,clip=,angle=270}}
\SetScale{0.283466457}
\SetWidth{1.763889}
{#4}
\end{picture}}
}
\newcommand{\psfigrol}[4]{%
\setcounter{pict@width}{1*\ratio{#2+\pict@scale/2}{\pict@scale}}
\setcounter{pict@height}{1*\ratio{#3+\pict@scale/2}{\pict@scale}}
\setlength{\unitlength}{\pict@scale}
\hbox{\begin{picture}(\thepict@width,\thepict@height)
\put(0,0){\psfig{figure=#1,width=#3,height=#2,clip=,angle=90}}
\SetScale{0.283466457}
\SetWidth{1.763889}
{#4}
\end{picture}}
}
\catcode`\@=12 % @ signs are no longer letters
\newlength\listtextwidth

\catcode`\@=11 % @ signs are now treated as letters
\newlength{\@tabfninsert}
\newlength{\@tabfnwidth}
\newcommand{\tabfootnote}[2]{%
  \setlength{\@tabfninsert}{0.8em}
  \setlength{\@tabfnwidth}{\textwidth}
  \addtolength{\@tabfnwidth}{-\@tabfninsert}
  \addtolength{\@tabfnwidth}{-0.4em}
  \noindent\makebox[\@tabfninsert][r]{\footnotesize$^{#1}$\hfil}\hfill%
  \parbox[t]{\@tabfnwidth}{\footnotesize #2\hfill}}
\catcode`\@=12 % @ signs are no longer letters

\def\citeCTD{{\cite{%
nim:a279:290,*npps:b32:181,*nim:a338:254%
}}\xspace}
\def\citeCAL{{\cite{%
nim:a309:77,*nim:a309:101,*nim:a321:356,*nim:a336:23%
}}\xspace}
\begin{document}
%------------------------------------------------------------------------------
%       Title sheet
%------------------------------------------------------------------------------
\prepnum{DESY--03--059}

\title{
Measurement of deeply virtual Compton scattering at HERA
}                                                       
                    
\author{ZEUS Collaboration}
\draftversion{4.5}
\date{April 2003}

\abstract{ The cross section for deeply virtual Compton scattering in
the reaction $ep \rightarrow e \gamma p$ has been measured with the
ZEUS detector at HERA using integrated luminosities of 95.0~pb$^{-1}$
of $e^+p$ and 16.7~pb$^{-1}$ of $e^-p$ collisions.  Differential cross
sections are presented as a function of the exchanged-photon
virtuality, $Q^2$, and the centre-of-mass energy, $W$, of the $\gamma
^*p$ system in the region $5 < Q^2 < 100\;{\rm GeV}^2$ and $40 < W <
140\;{\rm GeV}$. The measured cross sections rise steeply with
increasing $W$.  The measurements are compared to QCD-based
calculations.  }

\makezeustitle

\def\3{\ss}                                                                                        
\newcommand{\address}{ }                                                                           
\pagenumbering{Roman}                                                                              
                                    % this "%"s are for cosmetics only                             
%\begin{document}                                                                                   
                                                   %                                               
\begin{center}                                                                                     
{                      \Large  The ZEUS Collaboration              }                               
\end{center}                                                                                       
  S.~Chekanov,                                                                                     
  M.~Derrick,                                                                                      
  D.~Krakauer,                                                                                     
  J.H.~Loizides$^{   1}$,                                                                          
  S.~Magill,                                                                                       
  B.~Musgrave,                                                                                     
  J.~Repond,                                                                                       
  R.~Yoshida\\                                                                                     
 {\it Argonne National Laboratory, Argonne, Illinois 60439-4815}~$^{n}$                            
\par \filbreak                                                                                     
  M.C.K.~Mattingly \\                                                                              
 {\it Andrews University, Berrien Springs, Michigan 49104-0380}                                    
\par \filbreak                                                                                     
  P.~Antonioli,                                                                                    
  G.~Bari,                                                                                         
  M.~Basile,                                                                                       
  L.~Bellagamba,                                                                                   
  D.~Boscherini,                                                                                   
  A.~Bruni,                                                                                        
  G.~Bruni,                                                                                        
  G.~Cara~Romeo,                                                                                   
  L.~Cifarelli,                                                                                    
  F.~Cindolo,                                                                                      
  A.~Contin,                                                                                       
  M.~Corradi,                                                                                      
  S.~De~Pasquale,                                                                                  
  P.~Giusti,                                                                                       
  G.~Iacobucci,                                                                                    
  A.~Margotti,                                                                                     
  R.~Nania,                                                                                        
  F.~Palmonari,                                                                                    
  A.~Pesci,                                                                                        
  G.~Sartorelli,                                                                                   
  A.~Zichichi  \\                                                                                  
  {\it University and INFN Bologna, Bologna, Italy}~$^{e}$
% ...
\newpage                                                                                           
$^{\    1}$ also affiliated with University College London \\                                      
% ...
                                                           %                                       
                                                           %                                       
% \par         % if index listing & table fit to 1 page, put gap here                              
\newpage   % alternatively: go to newpage, if page is too small                                    
                                                           %                                       
% \institute_references_start    % do not touch or move this line !                                
                                                           %                                       
\begin{tabular}[h]{rp{14cm}}                                                                       
$^{a}$ &  supported by the Natural Sciences and Engineering Research                               
          Council of Canada (NSERC) \\                                                             
$^{b}$ &  supported by the German Federal Ministry for Education and                               
          Research (BMBF), under contract numbers HZ1GUA 2, HZ1GUB 0, HZ1PDA 5, HZ1VFA 5\\         
$^{c}$ &  supported by the MINERVA Gesellschaft f\"ur Forschung GmbH, the                          
          Israel Science Foundation, the U.S.-Israel Binational Science                            
          Foundation and the Benozyio Center                                                       
          for High Energy Physics\\                                                                
$^{d}$ &  supported by the German-Israeli Foundation and the Israel Science                        
          Foundation\\                                                                             
$^{e}$ &  supported by the Italian National Institute for Nuclear Physics (INFN) \\                
$^{f}$ &  supported by the Japanese Ministry of Education, Culture,                                
          Sports, Science and Technology (MEXT) and its grants for                                 
          Scientific Research\\                                                                    
$^{g}$ &  supported by the Korean Ministry of Education and Korea Science                          
          and Engineering Foundation\\                                                             
$^{h}$ &  supported by the Netherlands Foundation for Research on Matter (FOM)\\                   
$^{i}$ &  supported by the Polish State Committee for Scientific Research,                         
          grant no. 620/E-77/SPUB-M/DESY/P-03/DZ 247/2000-2002\\                                   
$^{j}$ &  partially supported by the German Federal Ministry for Education                         
          and Research (BMBF)\\                                                                    
$^{k}$ &  supported by the Fund for Fundamental Research of Russian Ministry                       
          for Science and Edu\-cation and by the German Federal Ministry for                       
          Education and Research (BMBF)\\                                                          
$^{l}$ &  supported by the Spanish Ministry of Education and Science                               
          through funds provided by CICYT\\                                                        
$^{m}$ &  supported by the Particle Physics and Astronomy Research Council, UK\\                   
$^{n}$ &  supported by the US Department of Energy\\                                               
$^{o}$ &  supported by the US National Science Foundation\\                                        
$^{p}$ &  supported by the Polish State Committee for Scientific Research,                         
          grant no. 112/E-356/SPUB-M/DESY/P-03/DZ 301/2000-2002, 2 P03B 13922\\                    
$^{q}$ &  supported by the Polish State Committee for Scientific Research,                         
          grant no. 115/E-343/SPUB-M/DESY/P-03/DZ 121/2001-2002, 2 P03B 07022\\                    
\end{tabular}                                                                                      
                                                           %                                       
% \institute_references_end     % do not touch or move this line !                                 
                                                           %                                       
%\end{document}                                                                                     

%------------------------------------------------------------------------------
%       Text
%------------------------------------------------------------------------------
\pagenumbering{arabic} 
\pagestyle{plain}
% ----------------------------------------------------------------------------
%       Introduction
% ----------------------------------------------------------------------------
%%%%%%%%%%%%%%%%%%%%%%%%%%%%%%%%%%%%%%%%%%%%%%%%%%%%%%%%%%%%%%%%%%%%%%%%%%
\section{Introduction}
\label{sec-int}
%%%%%%%%%%%%%%%%%%%%%%%%%%%%%%%%%%%%%%%%%%%%%%%%%%%%%%%%%%%%%%%%%%%%%%%%%%
%\input{yyintro}

\begin{figure}
\centering
\epsfig{file=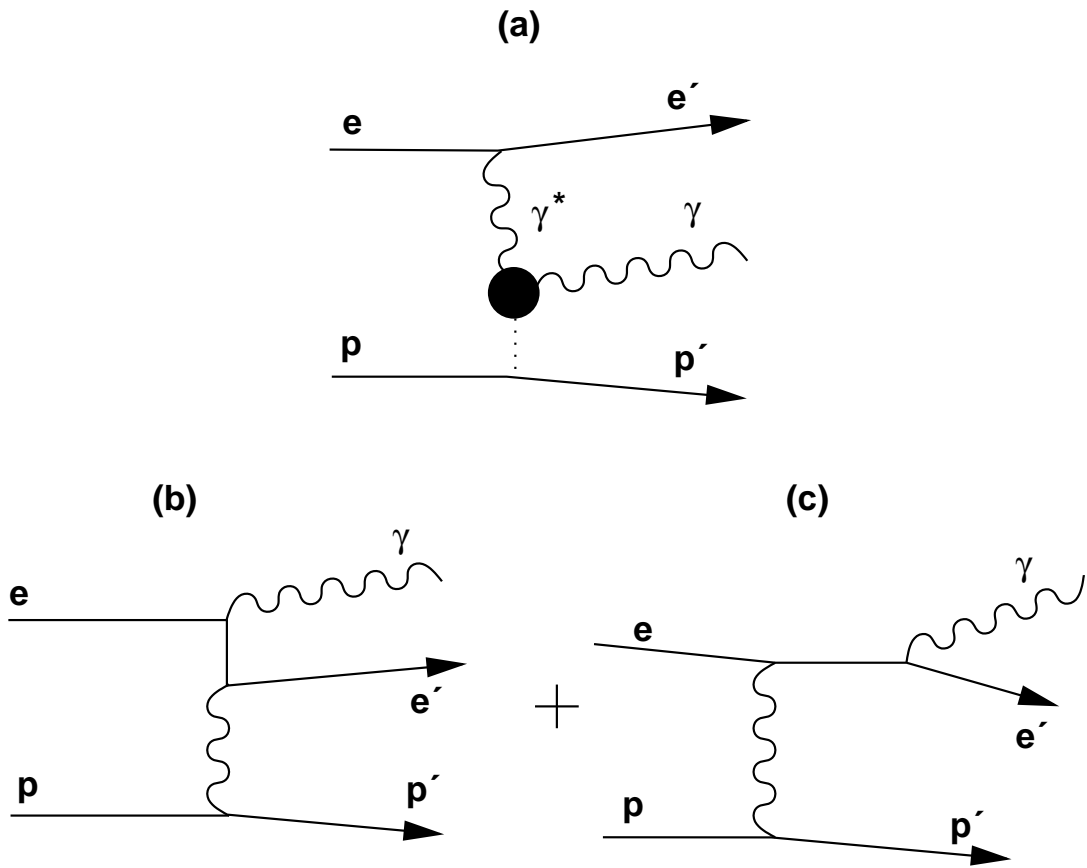,width=0.9\textwidth}
\caption{(a) Deeply virtual Compton scattering (DVCS) diagram,
  (b) and (c) Bethe-Heitler (BH) process.}
\label{fig-dvcsdiagram}
\end{figure}

This paper reports cross-section measurements for the exclusive
production of a real photon in diffractive $ep$ interactions,
$ep\rightarrow e\gamma p$, as shown in \fig{dvcsdiagram}a. This
exclusive process, known as deeply virtual Compton scattering
(DVCS)~\cite{tmp:19:14,*zfp:c12:263,*pr:d55:7114,pr:d58:114008,
pr:d59:74009,pr:d58:114001}, is calculable in perturbative QCD (pQCD),
when the virtuality, $Q^2$, of the exchanged photon is large.

The DVCS reaction can be regarded as the elastic scattering of the
virtual photon off the proton via a colourless exchange.  The pQCD
calculations assume that the exchange involves two partons, having
different longitudinal and transverse momenta, in a colourless
configuration. These unequal momenta are a consequence of the mass
difference between the incoming virtual photon and the outgoing real
photon. The DVCS cross section depends, therefore, on the generalised
parton distributions
(GPD)~\cite{Muller:1994,pr:d58:114001,prl:78:610,pr:d59:014029}, which carry
information about the wave function of the proton.  The cross section
at sufficiently large $Q^2$ is expected to rise steeply with
increasing $W$, the centre-of-mass energy of the virtual photon-proton
system, due to the fast rise of the parton densities in the proton
towards smaller $x$ values, where $x$ is the Bjorken scaling variable.

The initial and final states of the DVCS process are identical to
those of the purely electromagnetic Bethe-Heitler (BH) process
(Figs.~\ref{fig-dvcsdiagram}b,c). The interference between these two
processes in principle provides information about the real and
imaginary parts of the QCD scattering
amplitude~\cite{plb:460:417:1999,Belitsky:2001ns,Freund:2001hm}.
However, it is expected to be small in the kinematic region studied in
this paper~\cite{plb:460:417:1999,Belitsky:2001ns}.

The simplicity of the final state and the absence of complications due
to hadronisation mean that the QCD predictions are more reliable than
for many other exclusive final states. This reaction is one of the
theoretically best-understood exclusive QCD processes in $ep$
collisions. The first measurements of the DVCS process at high
$W$~\cite{misc:zeus:dvcs-talk,pl:b517:47} and its beam-spin asymmetry
in polarised $ep$ scattering at low
$W$~\cite{prl:87:182001,prl:87:182002} have recently become available.

In the analysis presented here, the dependence of the DVCS cross
section on $W$ and $Q^2$ is studied in the kinematic range
$5<Q^2<100\,\Gev^2$ and $40<W<140 \,\Gev$. The measurements are
integrated over $t$, the square of the four-momentum transfer at the
proton vertex.  The $e^{+}p$ cross sections are based on a ten-fold
increase in statistics over a previous HERA result~\cite{pl:b517:47},
permitting a study of the $W$ dependence of the cross section as well
as a significant extension of the $Q^2$ range probed.  This paper also
reports the first measurement of the $e^{-}p$ cross sections.

%%%%%%%%%%%%%%%%%%%%%%%%%%%%%%%%%%%%%%%%%%%%%%%%%%%%%%%%%%%%%%%%%%%%%%%%%%
\section{Experimental set-up}
\label{sec-zeus}
%%%%%%%%%%%%%%%%%%%%%%%%%%%%%%%%%%%%%%%%%%%%%%%%%%%%%%%%%%%%%%%%%%%%%%%%%%
The data were collected by the ZEUS detector at HERA during the 96-00
running periods. In 96-97, HERA collided \mbox{27.5 GeV} positrons
with \mbox{820 GeV} protons. In 98-00, the proton energy was increased
to 920 GeV and both positrons and electrons were collided. The
measurements for $e^+p$ ($e^-p$) interactions\footnote{Hereafter, both
$e^+$ and $e^-$ are referred to as electrons, unless explicitly stated
otherwise.} are based on an integrated luminosity of $95
\,\mathrm{pb^{-1}}$ ($17 \,\mathrm{pb^{-1}}$).

\Zdetdesc

%CTD
\Zctddesc\ZcoosysfnBeta

%CAL
\Zcaldesc

%Presampler
Presampler detectors~\cite{nim:a382:419,*magill:2000:bpres} are
mounted in front of the CAL. They consist of scintillator tiles
matching the calorimeter towers and measure signals from particle
showers created by interactions in the material lying between the
interaction point and the calorimeter. In this analysis, only the
information from the presampler in front of the RCAL was used to
correct the energy of the final-state particles.

%FPC
The forward plug calorimeter (FPC)~\cite{nim:a450:235} is a
lead-scintillator sandwich calorimeter with readout via
wavelength-shifter fibres. It was installed in 1998 in the $\rm{20
\times 20 \,cm^2}$ beam hole of the FCAL and has a small hole of
radius 3.15 cm in the centre to accommodate the beam pipe. It extends
the pseudorapidity coverage of the forward calorimeter from $\rm{\eta
< 4.0}$ to $\rm{\eta < 5.0}$.  The FPC information was used to remove
low-mass proton-dissociative events from the analysis.

%HES 
The hadron-electron separator (HES)~\cite{nim:a277:176} is installed
in the RCAL and FCAL. It consists of $3\times3\;\cm^2$ silicon diodes
placed at a longitudinal depth of three radiation lengths, which
corresponds to the approximate position of the maximum of the
electromagnetic shower in the CAL. The separation between electrons
and hadrons is based on the fact that the hadronic interaction length
is 20 times larger than the electromagnetic radiation length. In this
analysis, the fine segmentation of the RHES was used to improve the
position resolution for both scattered electrons and photons.

%SRTD
The small-angle rear tracking detector (SRTD)~\cite{nim:a401:63} is
attached to the front face of the RCAL ($Z=-148\;\rm{cm}$). The SRTD
consists of two planes of scintillator strips read out via optical
fibres and photomultiplier tubes. It covers the region $\rm{68 \times
  68 \,cm^2}$ in $X$ and $Y$ with the exclusion of a $\rm{8 \times 20
  \,cm^2}$ hole at the centre for the beam pipe. The SRTD provides a
transverse-position resolution of 3 mm and was used to measure the
positions of photons and electrons scattered at small angles relative
to the lepton beam direction.

%PRT1
The proton-remnant tagger (PRT1)~\cite{zfp:c75:421} consists of two
layers of scintillation counters located at $Z=5.15 \,\rm{m}$, and
covers the pseudorapidity range $\rm{4.3 < \eta < 5.8}$. It was used,
up to the end of the 1997 running period, to tag events in which the
proton diffractively dissociated.

%luminosity
The luminosity was determined from the rate of the bremsstrahlung
process $ep\rightarrow e\gamma p$, where the high-energy photon was
measured with a lead-scintillator
calorimeter~\cite{desy-92-066,*zp:c63:391,*acpp:b32:2025} located at
$Z=-107\,\rm{m}$.

%%%%%%%%%%%%%%%%%%%%%%%%%%%%%%%%%%%%%%%%%%%%%%%%%%%%%%%%%%%%%%%%
\section{Event selection}
\label{sec-strategy}
%%%%%%%%%%%%%%%%%%%%%%%%%%%%%%%%%%%%%%%%%%%%%%%%%%%%%%%%%%%%%%%%%
For the $Q^2$ range of this analysis, $Q^2>5 \gev^2$, and small $t$,
the signature of DVCS and BH events consists of a photon and a
scattered electron with balanced transverse momenta.  The scattered
proton stays in the beam pipe and remains undetected.

The events were selected online via a three-level trigger
system~\cite{zeus:1993:bluebook,Smith:1992}. The trigger selected
events with two isolated electromagnetic (EM) clusters in the EMC with
energy greater than 2 \Gev.  The events were selected offline by
requiring two EM clusters, the first in the RCAL with energy
$E_1>15\,\Gev$ and the second, with polar angle $0.6<\theta_2 <2.75
\,\mathrm{rad}$ ($1.2>\eta_2>-1.6$), either in the RCAL, with energy
$E_2>3 \,\Gev$, or in the BCAL, with energy $E_2>2.5 \,\Gev$. The
angular range of the second cluster corresponds to the region of high
efficiency for reconstruction of a track in the CTD. If a track was
found, it was required to match one of the EM clusters. Events with more than
one track were rejected.
To ensure full containment of the electromagnetic showers in the CAL,
events in which one of the clusters was located within 3 cm of the
beam hole were rejected.

The selection $40<E-p_Z<70\;\Gev$ was imposed, where $E$ is the total
energy and $p_Z$ the sum of $E\cos \theta$ over the whole CAL. This
requirement rejects photoproduction events and events in which a hard
photon is radiated from the incoming electron.

After these cuts, when the two EM clusters are ordered in energy such
that $E_1>E_2$, the kinematics ensure $\eta_1<\eta_2$. In the
following, the two clusters will be denoted as EM1 and EM2,
respectively.

For the 96-97~(98-00) running period, calorimeter cells not associated
with the two electromagnetic clusters were required to have energy
less than: 150~(200) MeV in the FEMC and 200~(300) MeV in the FHAC;
200~(350) MeV in the BEMC and 250~(350) MeV in the BHAC; 150~(150) MeV in
the REMC and 300~(300) MeV in the RHAC.  These thresholds were set to
be three standard deviations above the noise level of the CAL.
Moreover, for the 98-00 data sample, the energy measured in the FPC
was required to be less than 1~GeV. These elasticity requirements
reject most events in which the proton dissociates into a hadronic
system, $X$.

The events were subdivided into three samples, a first in which there
was no track associated with the EM2 cluster ($\gamma$ sample), a
second in which the track associated with EM2 had the same charge as
the beam electron ($e$ sample) and finally a third in which the track
associated with EM2 had the opposite charge to that of the beam
electron (\mbox{wrong-sign-$e$} sample). These samples are
interpreted as:

\begin{itemize}
\item $\gamma$ sample: EM2, with no track pointing to it, is the
  photon candidate and EM1 is the scattered-electron candidate. Both
  BH and DVCS processes contribute to this topology. The sample
  consisted of 3945 events.
  
\item $e$ sample: EM2, with the right-charge track pointing to it, is
  the scattered-electron candidate and EM1 is the photon candidate.
  This sample is dominated by the BH process. The contribution from
  DVCS is predicted to be negligible, due to the large $Q^2$ required
  for a large electron scattering angle. This sample contained 7059
  events.
  
\item \mbox{wrong-sign-$e$ sample}: EM2, with the wrong-charge-sign
  track pointing to it, may have originated from an $e^+ e^-$ final
  state accompanying the scattered electron, where one of the
  right-sign electrons escaped detection. This background sample is
  due to non-resonant $e^+ e^-$ production and to $J/\psi$ production
  and subsequent decay. Other sources are negligible, as will be
  discussed later. This sample consisted of 287 events.
\end{itemize}

The \mbox{wrong-sign-$e$ sample} was used to statistically subtract
the background contributions to the $e$ sample in each kinematic bin.
The background-subtracted $e$ sample was then used to investigate the
BH contribution to the $\gamma$ sample.

For the purposes of this analysis, the values of $Q^2$ and $W$ were
determined for each event, independently of its topology, under the
assumption that the EM1 cluster is the scattered electron. This
assumption is always valid for DVCS events for the $Q^2$ range
considered here.  The value of $Q^2$ was calculated using the electron
method~\cite{proc:hera:1991:23,*hoeger}, while $W$ was determined
using the double-angle method~\cite{proc:hera:1991:23,*hoeger}. No
explicit cut on $t$ was applied in the event selection.  Events for
which \mbox{$40<W<140 \,\Gev$} and \mbox{$5 < Q^2 < 100 \,\Gev^2$}
were retained.

%%%%%%%%%%%%%%%%%%%%%%%%%%%%%%%%%%%%%%%%%%%%%%%%%%%%%%%%%%%%%%%%%%%%%%%%%%
\section{Monte Carlo simulations}
\label{sec-mc}
%%%%%%%%%%%%%%%%%%%%%%%%%%%%%%%%%%%%%%%%%%%%%%%%%%%%%%%%%%%%%%%%%%%%%%%%%%
The acceptance and the detector response were determined using Monte
Carlo (MC) simulations. The detector was simulated in detail using a
program based on GEANT~3.13~\cite{misc:cern:geant3}. All of the
simulated events were processed through the same reconstruction and
analysis chain as the data. 
%The simulation of the trigger was adjusted
%to reproduce the trigger efficiency determined from the
%data~\cite{thesis:bold:2003}.

A MC generator, GenDVCS~\cite{misc:gendvcs} based on a model by
Frankfurt, Freund and Strikman (FFS)~\cite{pr:d58:114001}, was used to
simulate the elastic DVCS process.  In the FFS calculation, the DVCS
cross section, integrated over the angle between the $e$ and $p$
scattering planes, is related to the inclusive structure-function
$F_2$ through
\begin{displaymath}
\frac{d^3\sigma_{\rm{DVCS}}^{ep\rightarrow e \gamma p}}{dx dQ^2 dt}=
\frac{\pi^2\alpha^3
}{2xR^2Q^6} \left[1+(1-y)^2\right]e^{-b|t|}F_2^2(x,Q^2)(1+\rho^2)\ ,
%\label{eq:xsec_dvcs}
\end{displaymath}
where $x \simeq Q^2/(Q^2+W^2)$ is the Bjorken scaling variable, $b$ is
the exponential slope of the $t$ dependence and $y$ is the fraction of
the electron energy transferred to the proton in its rest frame. The
ratio $R=\Im m \mathcal{A}(\gamma^\ast p\rightarrow
\gamma^\ast p)|_{t=0}/ \Im m \mathcal{A}(\gamma^\ast p\rightarrow
\gamma p)|_{t=0}$ accounts for the non-forward character of the DVCS
process and is directly related to a ratio of the GPD to the parton
distribution functions~\cite{misc:fms:gpd} and $\rho$ is the ratio of
the real to imaginary part of the amplitude, $\rho=\Re
e\mathcal{A}(\gamma^\ast p\rightarrow \gamma p)|_{t=0} /\Im
m\mathcal{A}(\gamma^\ast p\rightarrow \gamma p)| _{t=0}$.

The value of $R$, calculated using the leading-order (LO) QCD evolution
of the GPD, is about 0.55, with little dependence on $x$ or
$Q^2$~\cite{pr:d58:114001}. For simplicity, the $R$ parameter in the
MC generator was set to a constant value of $R=0.55$.

%The value of $\rho$ was determined from the $\ln (1/x)$ derivative of
%the imaginary part of the DVCS amplitude. 
In GenDVCS, the ALLM97~\cite{hep-ph-9712415} parameterisation of the
$F_2$ structure function of the proton was used as input. In this
empirical fit to $\gamma^\ast p$ total-cross-section data, the value
of $\rho$ was parameterised as $\rho=\frac{\pi}{2}(0.176+0.033\ln
Q^2)$, where $Q^2$ is in $\gev^2$~\cite{misc:gendvcs}.

In the FFS model, the $t$ dependence is assumed to factorise, with the
slope parameter, $b$, depending on both $W$ and $Q^2$. The value of
$b$ is expected to decrease with $Q^2$ and, even at high $Q^2$ and at
very small $x$, is expected to increase with $W$. While this
dependence is important for the normalisation of the calculated DVCS
cross section, it does not affect the acceptance corrections.  In the
MC simulation $b$ was assumed to be constant and was set to 4.5
$\rm{GeV^{-2}}$.

For proper treatment of radiative effects, the GenDVCS generator was
interfaced to HERACLES~4.6~\cite{spi:www:heracles}, which includes
corrections for initial- and final-state photon emission from the
electron line, as well as vertex and propagator corrections.

The elastic and inelastic BH processes, $ep\rightarrow e\gamma p$ and
$ep\rightarrow e\gamma X$, and the QED di-lepton production,
$ep\rightarrow ee^+e^-p$, were simulated using the
GRAPE-Compton\footnote{Hereafter, the GRAPE-Compton generator is
referred as GRAPE}~\cite{cpc:136:126} and
GRAPE-Dilepton~\cite{cpc:136:126} generators, respectively. These two
MC programs are based on the automatic system
GRACE~\cite{misc:kek:grape-compton} for calculating Feynman diagrams.
The GRAPE generator gave identical results to the Compton
2.0~\cite{proc:hera:1991:1468} generator for the elastic BH process.
The GRAPE program was used because it simulates the hadronic final
state for the inelastic BH process.

Additional samples were generated using the diffractive
RAPGAP~\cite{cpc:86:147} and non-diffractive
DJANGOH~\cite{spi:www:djangoh11} generators in order to study possible
backgrounds from low-multiplicity DIS events. A possible contribution
from vector-meson electroproduction was simulated by the
ZEUSVM~\cite{thesis:muchorowski:1998} MC generator interfaced to
HERACLES.

%%%%%%%%%%%%%%%%%%%%%%%%%%%%%%%%%%%%%%%%%%%%%%%%%%%%%%%%%%%%%%%%
\section{DVCS-signal extraction}
\label{sec-analysis}
%%%%%%%%%%%%%%%%%%%%%%%%%%%%%%%%%%%%%%%%%%%%%%%%%%%%%%%%%%%%%%%%%

In the kinematic region of this analysis the interference between the
DVCS and BH amplitudes is very small when the cross section is
integrated over the angle between the $e$ and $p$ scattering
planes~\cite{plb:460:417:1999,Belitsky:2001ns}.  Thus the cross
section for exclusive production of real photons may be treated as a
simple sum over the contributions from the DVCS and electromagnetic BH
processes. The latter can, therefore, be subtracted and the DVCS cross
section determined.

The BH process was studied using the $e$ sample which, according to
the MC predictions, consists almost solely of BH candidates. A
background contribution to the BH events in this sample of about 4\%,
originating from deep inelastic exclusive $e^+e^-$ production, where
one of three final-state leptons escapes detection, was estimated from
the \mbox{wrong-sign-$e$ sample} and MC simulations, and was
statistically subtracted. According to the MC simulations, 75\% of
this background consists of non-resonant di-lepton production and 25\%
of exclusive $J/\psi$ production with subsequent $e^+e^-$ decay.  The
normalisation of the MC samples was determined from the
\mbox{wrong-sign-$e$ sample}.  The diffractive electroproduction of
$\rho$, $\omega$ and $\phi$ mesons, in which one of the decay charged
particles was misidentified as an electron in the CAL, the other was
undetected, and the electron scattered into the RCAL was taken to
be the photon, was negligible.

The expectation for the inelastic BH contribution to the $e$ sample is
subject to uncertainties coming from the dependence of the selection
efficiency on the mass of the hadronic final-state system, $X$.  This
inelastic contribution was estimated from the data as $(17.8 \pm
1.2)$\%, for the 96-97 data sample, using the fraction of events
tagged in the PRT1 and $(10.5 \pm 1.0)$\%, for the 98-00 data sample,
using events with more than 1~GeV of energy in the FPC, obtained
releasing the elasticity cut. The uncertainties are statistical. No
attempt was made to quantify the systematic uncertainty since there is
little sensitivity to the cross section. The difference in the
measured fraction of proton-dissociative events is due to differences
in the detector configuration between the two running periods.

After the subtraction of the dilepton and $J/\psi$ backgrounds, for
the 96-97 (99-00) $e^+p$ samples, the number of remaining BH events in
the $e$ sample was 2523 (3289), while the expected number from the
GRAPE simulation was 2601 (3358). The absolute expectation of the
GRAPE simulation reproduced the number of BH data events to within
$(3\pm 5)\%$ for 96-97 and $(2 \pm 4)\%$ for 99-00, where
uncertainties include the statistical uncertainty as well as the
uncertainties due to the trigger efficiency and the estimation of the
inelastic BH contribution.

\begin{figure}
\centering
\epsfig{file=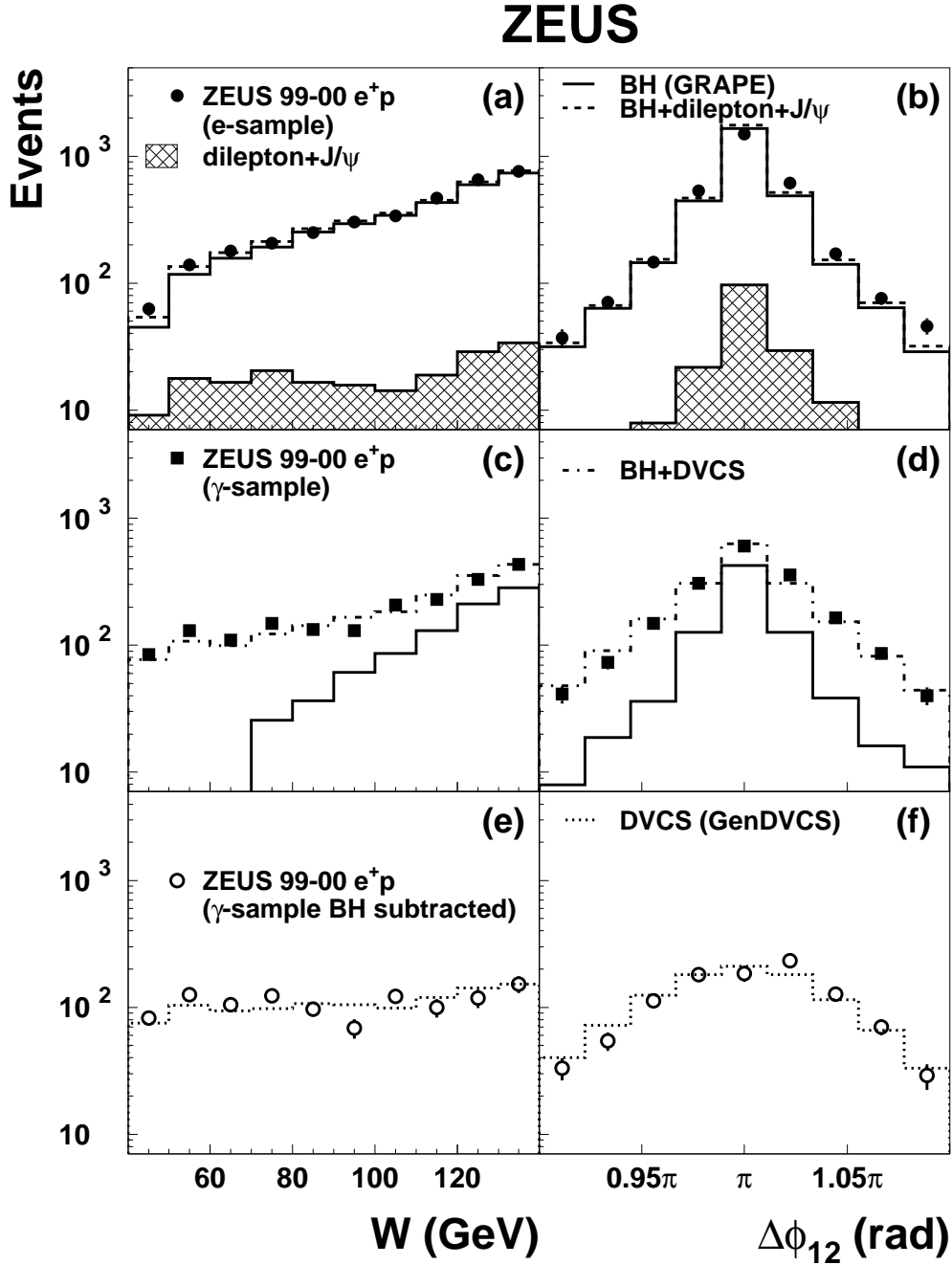,width=0.9\textwidth,bb=15 0 480 580}
\caption{Comparison of the $W$ and the $\Delta\phi_{12}$ distributions
for the data from 99-00 $e^+p$ running period and corresponding MC
samples as described in the figure; (a) and (b) $e$ sample compared to
the sum of the BH, dilepton and $J/\psi$ MC simulations, with the sum
of the latter two also shown separately; (c) and (d) $\gamma$ sample
compared to the expectation of the BH MC simulation, normalised to the
data in the $e$ sample after background subtraction; (e) and (f)
$\gamma$ sample after subtracting the BH expectation compared to
GenDVCS, where the absolute normalisation of the latter is increased
by 17.5\% to account for the proton-dissociative component in the DVCS
data.}
\label{fig:dvcs2}
\end{figure}

For further analysis, the GRAPE MC sample was normalised to the BH
events in the data. The comparison between the $e$ sample and the sum
of the GRAPE BH, dilepton and $J/\psi$ MC samples for the 99-00 $e^+p$
running period is shown in Fig.~\ref{fig:dvcs2}a versus $W$ and in
Fig.~\ref{fig:dvcs2}b versus the difference in azimuthal angles of the
EM1 and the EM2, $\Delta\phi_{12}$.  Good agreement is observed.

The properties of the $\gamma$ sample were then compared to the
expectations of the normalised GRAPE MC sample. As an example, the
comparison of the $W$ and $\Delta \phi_{12}$ distributions for the
99-00 $e^+p$ running period is shown in Figs.~\ref{fig:dvcs2}c
and~\ref{fig:dvcs2}d, respectively.  An excess of events over the
expectations of the GRAPE simulation is observed.  Moreover, the data
distributions differ from those expected for the BH process.  The $W$
distribution of the BH sample peaks at large $W$, while that of the
data is more evenly distributed. The $\Delta \phi_{12}$ distribution
of the BH sample is also narrower than that of the data.

The $W$ and the $\Delta \phi_{12}$ distributions of the data, after
subtracting the BH contribution using the renormalised GRAPE MC
sample, are shown in Figs.~\ref{fig:dvcs2}e and~\ref{fig:dvcs2}f,
respectively. This sample includes events in which the proton
dissociated into a hadronic final state with low mass.
 
The previous ZEUS measurements of elastic vector meson
production~\cite{epj:c6:603,*epj:c14:213,hep-ex-0201043} support, within relatively large
uncertainties, the assumption that the fraction of proton dissociative
events, $f_{\rm{p-diss}}$, in diffractive interaction is process independent.
Therefore, in this analysis, the values of $f_{\rm{p-diss}}$
determined from the measurements of the diffractive J/psi
photoproduction~\cite{hep-ex-0201043} are used
%To correct for
%this contribution, it was assumed that the fraction of
%proton-dissociative events, $f_{\rm{p-diss}}$, in diffractive
%interactions is process independent and also independent of $W$ and
%$Q^2$, so that the values determined from diffractive $J/\psi$
%photoproduction~\cite{hep-ex-0201043} can be used:
\begin{center}
\begin{tabular}{cccc}
$f_{\rm{p-diss}}$&=&$\rm{22.0\pm2.0(stat.)\pm2.0(syst.)\%}$ & for the 96-97 data; \\
$f_{\rm{p-diss}}$&=&$\rm{17.5\pm 1.3(stat.)^{+3.7}_{-3.2}(syst.)\%}$ & for the
98-00 data.\\
\end{tabular}
\end{center}
The above fractions are consistent, within large uncertainties, with
those estimated using the events in the $\gamma$ sample either tagged
by the PRT1 or the FPC, after subtracting the inelastic BH
contributions.

Other possible sources of contamination were investigated. Due to the
relatively high $Q^2$ of the present data set, the contamination from
production of light vector mesons, such as $\omega$ or $\phi$,
decaying through channels containing photons in the final state is
below 1\% and was neglected. A possible contribution to the $\gamma$
sample from low-multiplicity processes such as $ep\rightarrow
e\pi^0p$, $ep\rightarrow e\pi^0\pi^0p$, and $ep\rightarrow e\pi^0\eta
p$, where the $\pi^0$ or $\eta$ fakes a photon signal, was also
investigated. The number of candidate events found in the RAPGAP and
DJANGOH samples was reweighted to reflect the cross sections obtained
by extrapolating low-$W$
measurements~\cite{pr:d7:1937,alekhin:xsec:cern:hera}.  Their
contribution to the DVCS sample is negligible.

%The $W$ and $\Delta \phi_{12}$ distributions of the DVCS data sample,
%obtained by subtracting the BH MC expectations from the $\gamma$
%sample, are displayed in Fig.~\ref{fig:dvcs2}e) and~\ref{fig:dvcs2}f),
%respectively. The expected number of events is corrected to account
%for the contribution of the proton-dissociative events.  

The data are compared to the absolute expectations of GenDVCS in
Figs.~\ref{fig:dvcs2}e and~\ref{fig:dvcs2}f.  The best agreement in
normalisation between the data and the MC simulation is achieved when
the normalisation of the latter is decreased by 10\%. This was
obtained by increasing the value of $b$ from $4.5$ to $4.9\;\Gev^{-2}$
(see Section~\ref{sec-mc}).  Overall, good agreement between the data
and the MC simulation is found, demonstrating that the excess of
photon candidates over the expectation of BH is due to DVCS.

%%%%%%%%%%%%%%%%%%%%%%%%%%%%%%%%%%%%%%%%%%%%%%%%%%%%%%%%%%%%%%%%
\section{Cross-section determination}
\label{sec-xsec}
%%%%%%%%%%%%%%%%%%%%%%%%%%%%%%%%%%%%%%%%%%%%%%%%%%%%%%%%%%%%%%%%%
The $\gamma^\ast p$ cross section for the DVCS process as a function
of $W$ and $Q^2$ was evaluated using the expression

\begin{displaymath}
\sigma(\gamma^\ast p \rightarrow \gamma p)(W_i, Q^2_i)=\frac{(N_i^{\rm{obs}}-N_i^{\rm{BH}})\cdot(1-f_{\rm{p-diss}})}{N_i^{\rm{MC}}}\cdot \sigma ^{\rm{FFS}}(\gamma^\ast p \rightarrow \gamma p)(W_i, Q^2_i),
\label{eq-ffs}
\end{displaymath}

where $N^{\rm{obs}}_i$ is the total number of data events in the
$\gamma$ sample in bin $i$ in $W$ and $Q^2$, $N^{\rm{BH}}_i$ denotes
the number of BH events in the $\gamma$ sample in that bin, determined
from the renormalised GRAPE sample, and $N^{\rm{MC}}_i$ is the number
of events expected in the $\gamma$ sample from GenDVCS for the
luminosity of the data. The factor $f_{\rm{p-diss}}$ is the fraction
of the proton-dissociative DVCS events in the data, $\sigma
^{\rm{FFS}}(\gamma^\ast p \rightarrow \gamma p)$ is the $\gamma ^*p$
cross section computed according to the FFS expression, and $W_i$ and
$Q^2_i$ are the values of $W$ and $Q^2$ where the cross section is
evaluated.

The $\gamma^\ast p$ cross sections have been computed in the ranges
$5<Q^2<100 \gev^2$ and $40<W<140 \gev$, separately for the 96-97, 98-99 and
99-00 data periods and then combined for the positron samples (96-97
and 99-00).  
%The mean values of $W$ and $Q^2$ for all the three
%samples, obtained from GenDVCS, are close to $\langle W\rangle=89
%\;\Gev$ and $\langle Q^2\rangle =9.6\;\Gev^2$, fairly independent of
%the sample.  
Tables~\ref{tab:tab1} -~\ref{tab:tab4} list the $\gamma
^*p\rightarrow \gamma p$ cross-section values.

%%%%%%%%%%%%%%%%%%%%%%%%%%%%%%%%%%%%%%%%%%%%%%%%%%%%%%%%%%%%%%%%
\section{Systematic uncertainties}
\label{sec-syst}
%%%%%%%%%%%%%%%%%%%%%%%%%%%%%%%%%%%%%%%%%%%%%%%%%%%%%%%%%%%%%%%%%
The systematic uncertainties of the measured cross sections were
determined by changing the selection cuts or the analysis procedure in
turn and repeating the extraction of the cross sections.  The
following systematic studies have been carried out:

\begin{itemize}
  
\item all the selection cuts discussed in Section~\ref{sec-strategy}
  were shifted according to the resolutions of the corresponding
  variables. The most significant contributions came from varying the
  lower $Q^2$ cut.  The average change in the cross section due to
  this cut was $\pm 2\%$. The largest change in the cross section,
  $\pm$10\%, was found in the highest-$W$ bin, while it was $\pm 4\%$
  in the lowest-$Q^2$ bin;

\item the elasticity cut was changed by $\pm30$~MeV in the EMC and
$\pm 50$~MeV in the HAC. The average change in the cross section was
$\pm$2\% in all bins of $Q^2$ and $W$, while the largest change in the
cross section, observed when the cut was lowered, was $-$4\% in the
lowest-$W$ bins and $-$4\% in the lowest-$Q^2$ bin;

%\item the fraction of the inelastic component in the BH events was
%varied in the range $(17.8 \pm 1.2)\%$ for 96-97 data and $(10.5 \pm
%1.0)\%$ for the 98-00 data sets, leading to a change in the
%cross section of about $\pm$1\%, evenly distributed over the bins. 
 
\item the trigger efficiency was varied within its statistical
  uncertainty. This resulted in average changes of the cross
  section of about $\pm$2\%. The biggest variation of the cross
  section of $\pm$3\% was observed in the lowest-$Q^2$ bin and in the
  two highest-$W$ bins;
  
\item the electromagnetic energy scale was varied within its
  uncertainty of $1.5\%$ for the EM2 (low energy) and of $1\%$ for the
  EM1 (high energy), resulting in a $\pm$3\% average change of the
  cross section in both $Q^2$ and $W$. The largest change was $\pm$3\%
  for the lowest-$Q^2$ bins and $\pm$5\% for the highest-$W$ bin;
  
\item in GenDVCS, the $Q^2$ dependence was modified by introducing a
  $Q^2$-dependent $t$ slope using the formula $b=8(1-0.15\ln
  (Q^2/2))$~GeV$^{-2}$ (see Section~\ref{sec-models}). The average
  change in the cross section was $\pm$1\%, with the largest variation
  of $\pm$3\% in the highest-$Q^2$ bin.

\end{itemize}

The uncertainty on the proton-dissociative contribution,
$f_{\rm{p-diss}}$, leads to an overall normalisation uncertainty of
$\pm4.0\%$ and $\pm3.5\%$ for the $e^-p$ and $e^+p$ data,
respectively.

The systematic uncertainties typically are small compared to the
statistical uncertainties.  The individual systematic uncertainties,
including that due to $f_{\rm{p-diss}}$, were added in quadrature
separately for the positive and negative deviations from the nominal
cross-section values to obtain the total systematic uncertainties
listed in Tables~\ref{tab:tab1} -~\ref{tab:tab4}.  An overall
normalisation uncertainty in the luminosity determination of
$\pm$1.8\% and $\pm$2.0\% for the $e^-p$ and $e^+p$ data,
respectively, was not included because it was small with respect to
the above contributions.

%%%%%%%%%%%%%%%%%%%%%%%%%%%%%%%%%%%%%%%%%%%%%%%%%%%%%%%%%%%%%%%%
\section{Results}
\label{sec-results}
%%%%%%%%%%%%%%%%%%%%%%%%%%%%%%%%%%%%%%%%%%%%%%%%%%%%%%%%%%%%%%%%%

\begin{figure}
\centering
\epsfig{file=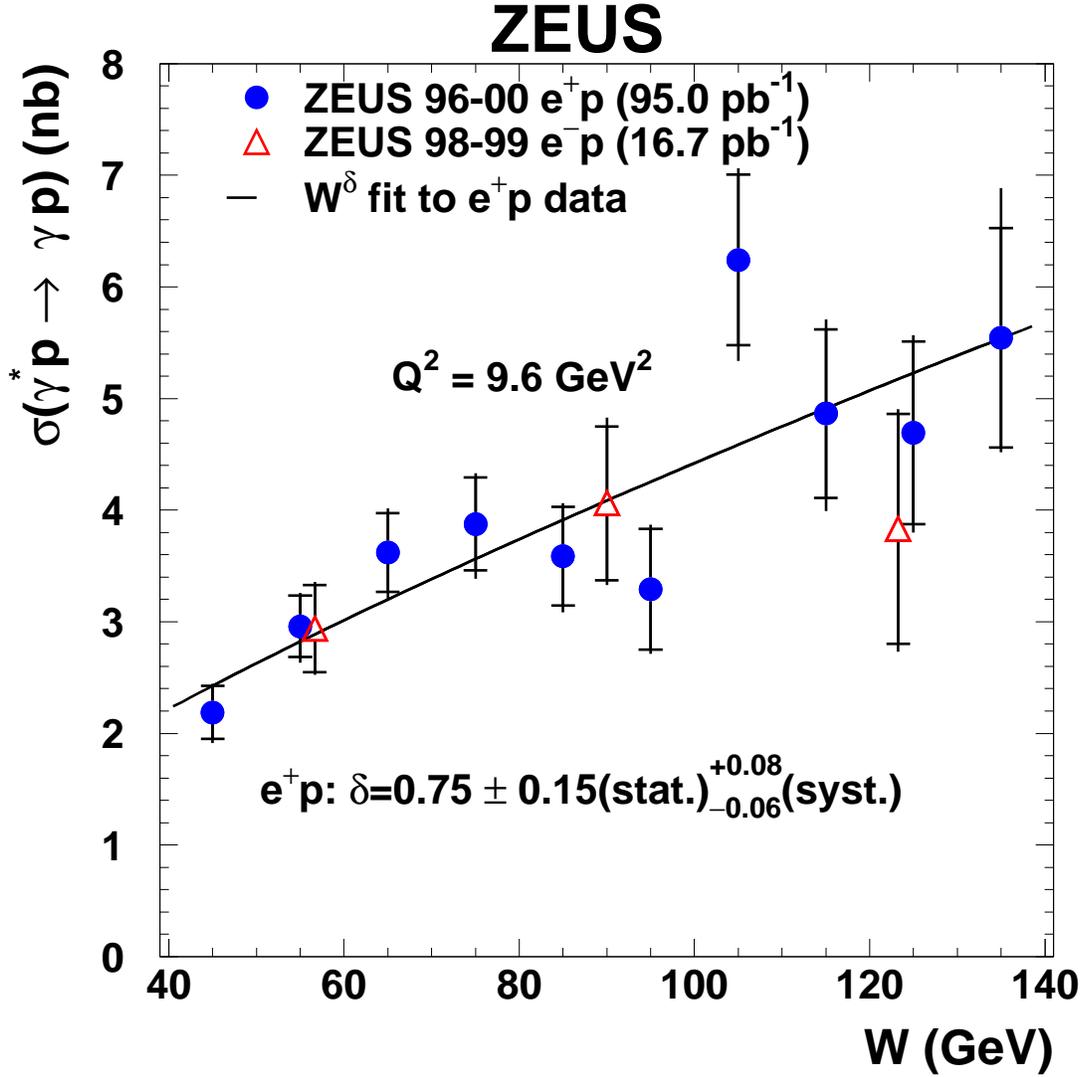,width=0.99\textwidth}
\caption{The DVCS cross section, $\sigma(\gamma ^*p\rightarrow \gamma
p)$, as a function of $W$ for an average $Q^2=9.6\;\Gev^2$, separately
for $e^+p$ data (dots) and $e^-p$ data (triangles).  The solid line is
the result of a fit of the form $\sigma_{\rm{DVCS}} \propto W^\delta$
to the positron data.  The error bars denote the statistical
uncertainty (inner) and the quadratic sum of the statistical and the
systematic uncertainties (outer).
\label{fig:e_p_w}}
\end{figure}

The $W$ dependence of the DVCS cross section,
$\sigma_{\mathrm{DVCS}}=\sigma(\gamma^\ast p \rightarrow \gamma p)$,
for $Q^2=9.6\,\Gev^2$ is shown in Fig. \ref{fig:e_p_w}, separately for
$e^+p$ and $e^-p$ interactions.  Due to the limited statistics, the
$e^-$ sample is only shown in three $W$ bins. There is agreement
between the two samples.

A fit of the form $\sigma_{\rm{DVCS}} \propto W^\delta$ was performed
separately for the positron and electron data. For the $e^+p$ data,
the value $\delta =0.75\pm
0.15(\mathrm{stat.})^{+0.08}_{-0.06}(\mathrm{syst.})$ is comparable to
that determined for $J/\psi$ electroproduction~\cite{hep-ex-0201043}.
This steep rise in the cross section is a strong indication of the
presence of a hard underlying process. The same fit to the $e^-p$ data
yields
$\delta=0.45\pm0.36(\mathrm{stat.})^{+0.08}_{-0.07}(\mathrm{syst.})$,
which is compatible with the $e^+p$ result.

\begin{figure}
\centering
\epsfig{file=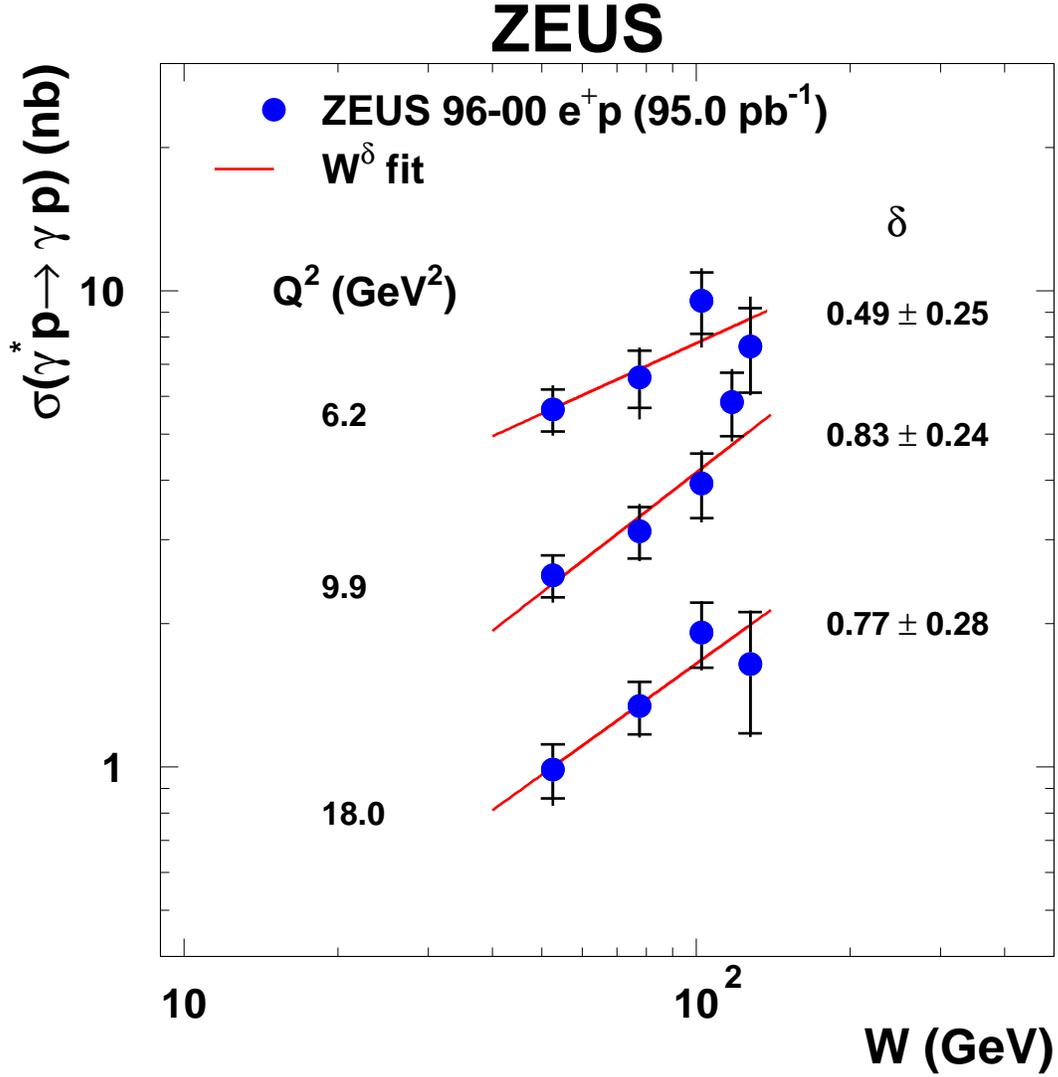,width=0.99\textwidth}
\caption{The DVCS cross section, $\sigma(\gamma ^*p\rightarrow \gamma
p)$, as a function of $~W$ for three $Q^2$ values for $e^+p$ data as
denoted in the figure. The corresponding ranges in $Q^2$ are listed in
Table~\ref{tab:tab4}.  The solid line is the result of a fit of the
form $\sigma_{\rm{DVCS}} \propto W^\delta$. The values of $\delta$ and
their statistical uncertainties are given in the figure.  The last
data point for $Q^2=9.9\gev^2$ is displaced horizontally for ease of
visibility. The error bars denote the statistical uncertainty (inner)
and the quadratic sum of the statistical and the systematic
uncertainties (outer).}
\label{fig:delta_w}
\end{figure}

The positron sample has been further subdivided into three $Q^2$
ranges. The $W$ dependence of $\sigma_{\rm{DVCS}}$ in these three
$Q^2$ bins is presented in Fig.~\ref{fig:delta_w}.
%There is an indication that the value of $\delta$ at $Q^2=6.2 \gev^2$ 
%is lower than the values of $\delta$ obtained at the two higher $Q^2$ bins.  
The results are compatible with no dependence of $\delta$ on $Q^2$
although also with the increase with $Q^2$ observed in exclusive
production of light vector mesons~\cite{epj:c13:371,epj:c6:603}.

\begin{figure}
\centering
\epsfig{file=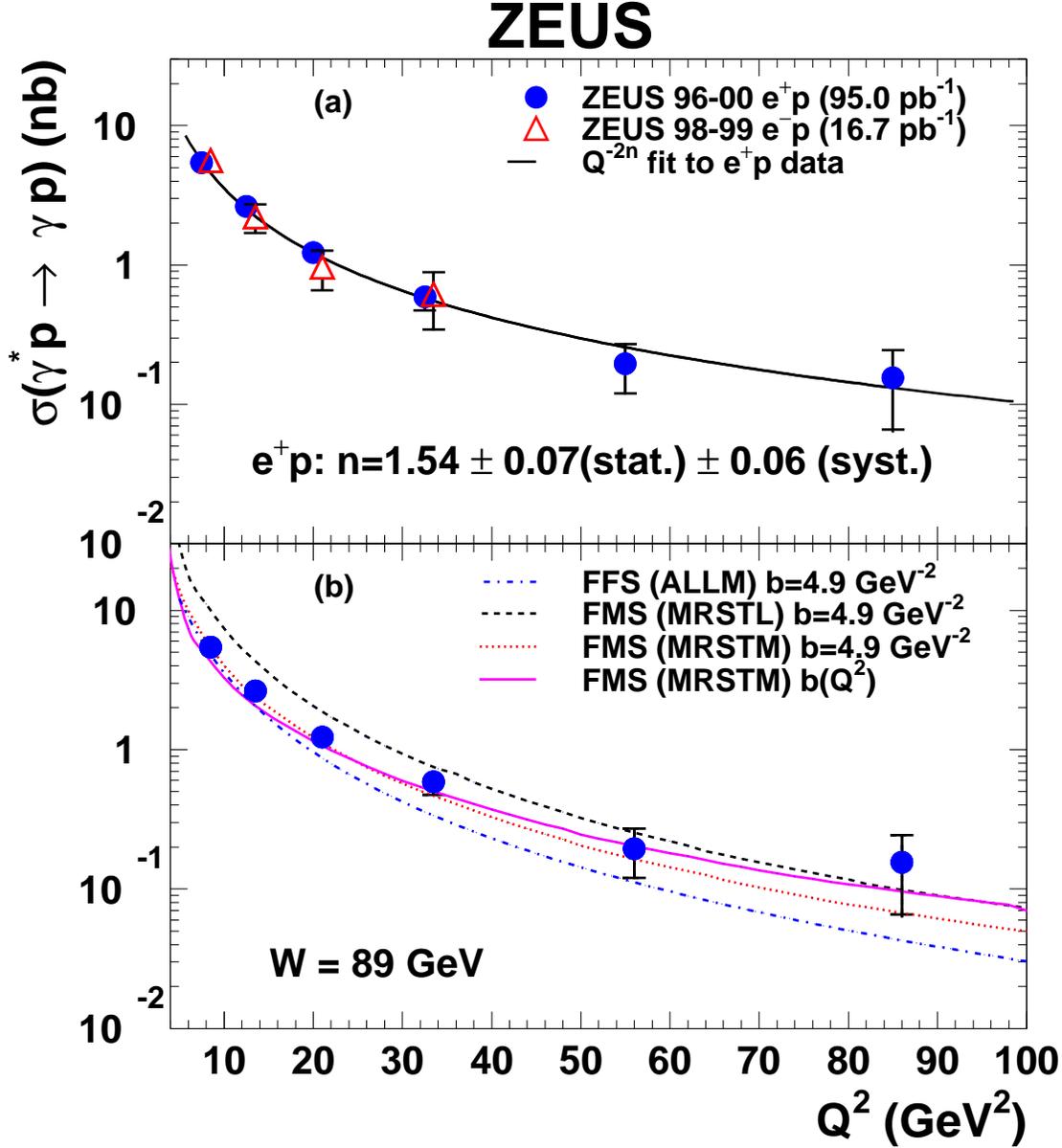,width=0.99\textwidth}
\caption{(a) The DVCS cross section, $\sigma(\gamma ^*p\rightarrow
\gamma p)$, as a function of $Q^2$ for $W=89\;\Gev$, separately for
$e^+p$ data (dots) and $e^-p$ data (triangles).  The solid line is the
result of a fit of the form $\sigma_{\rm{DVCS}} \propto Q^{-2n}$ to
the positron data.  The $e^-p$ data points are displaced horizontally
for ease of visibility; (b) $\sigma(\gamma ^*p\rightarrow \gamma p)$
as a function of $Q^2$ compared to the GPD-based theoretical
predictions of FFS and FMS, where MRSTL(M) indicates the LO (NLO)
parameterisation of PDF.  The MRSTM expectations are also shown for
the $Q^2$-dependent $b$ values described in the text.  The error bars
denote the statistical uncertainty (inner) and the quadratic sum of
the statistical and the systematic uncertainties (outer).
\label{fig:q2_models}}
\end{figure}

The $Q^2$ dependence of $\sigma_{\rm{DVCS}}$, for $W =89 \,\Gev$, is shown
in Fig. \ref{fig:q2_models}a, again separately for $e^+p$ and $e^-p$
interactions. There is no significant cross-section difference between
the $e^+$ and $e^-$ data, which is consistent with the assumption that
the present measurement is insensitive to the interference term.

A fit of the form $Q^{-2n}$ to the $e^+p$ data gives a value of
$n=1.54 \pm 0.07(\mathrm{stat.})\pm 0.06(\mathrm{syst.})$.  This value
is lower than $n\simeq 2$ which is characteristic of exclusive
vector-meson production~\cite{epj:c6:603,pl:b483:360}. The fit to the
$e^-p$ data gives $n=1.69 \pm
0.21(\mathrm{stat.})^{+0.09}_{-0.06}(\mathrm{syst.})$.

%%%%%%%%%%%%%%%%%%%%%%%%%%%%%%%%%%%%%%%%%%%%%%%%%%%%%%%%%%%%%%%%
\section{Comparison with models}
\label{sec-models}
%%%%%%%%%%%%%%%%%%%%%%%%%%%%%%%%%%%%%%%%%%%%%%%%%%%%%%%%%%%%%%%%%

In the presence of a hard scale ($Q^2 > > \Lambda^2_{\rm{QCD}}$), the
DVCS amplitude factorises into a hard-scattering coefficient,
calculable in pQCD, and a soft part which can be absorbed in the
GPD~\cite{pr:d59:74009}. The kernels of the evolution equations for
the GPD are known to next-to-leading order
(NLO)~\cite{Belitsky:1999gu,epj:c23:651,*prd:d65:091901} 
and the GPD can thus be
evaluated at all $Q^2$ given an input at some starting
scale. Measurements of the DVCS cross section are an essential
ingredient in modelling the input
GPD~\cite{epj:c23:651,*prd:d65:091901,misc:fms:gpd}.

Freund, McDermott and Strikman (FMS)~\cite{misc:fms:gpd} have made an
attempt to model the GPD based on DVCS
data~\cite{pl:b517:47,prl:87:182001,prl:87:182002}.  A comparison of
$\sigma_{\rm{DVCS}}$ as a function of $Q^2$ for fixed $W$ with the
predictions based on the MRST parameterisation of the parton
distribution functions (PDF)~\cite{hep-ph-0106075,npps:79:105} is
shown in Fig.~\ref{fig:q2_models}b. Three FMS curves are shown. Two
curves show the results of modelling based on LO (MRSTL) and NLO
(MRSTM) parton distribution functions. The latter leads to predictions
closer to the data. In this comparison, a fixed value of the $t$
slope, $b=4.9$~GeV$^{-2}$, was assumed.  The third curve, shown in the
figure, corresponds to predictions based on MRSTM, assuming a
$Q^2$-dependent $b$ value. The best agreement between the data and the
predictions is achieved using $b=8(1-0.15\ln (Q^2/2))$~GeV$^{-2}$, a
parameterisation obtained by Freund, McDermott and
Strikman~\cite{misc:fms:gpd} from a fit to a preliminary version of
the present data.  Similar conclusions are reached when the CTEQ6
parameterisations~\cite{pr:d51:4763,pl:b304:159,epj:c12:375} are used
(not shown).

The data are also compared to the expectations of FFS (see
Section~\ref{sec-mc}), again assuming $b=4.9\;\Gev^{-2}$. For
$Q^2>20\gev^2$, the $e^+p$ data lie significantly above the
prediction.

\begin{figure}
\centering
\epsfig{file=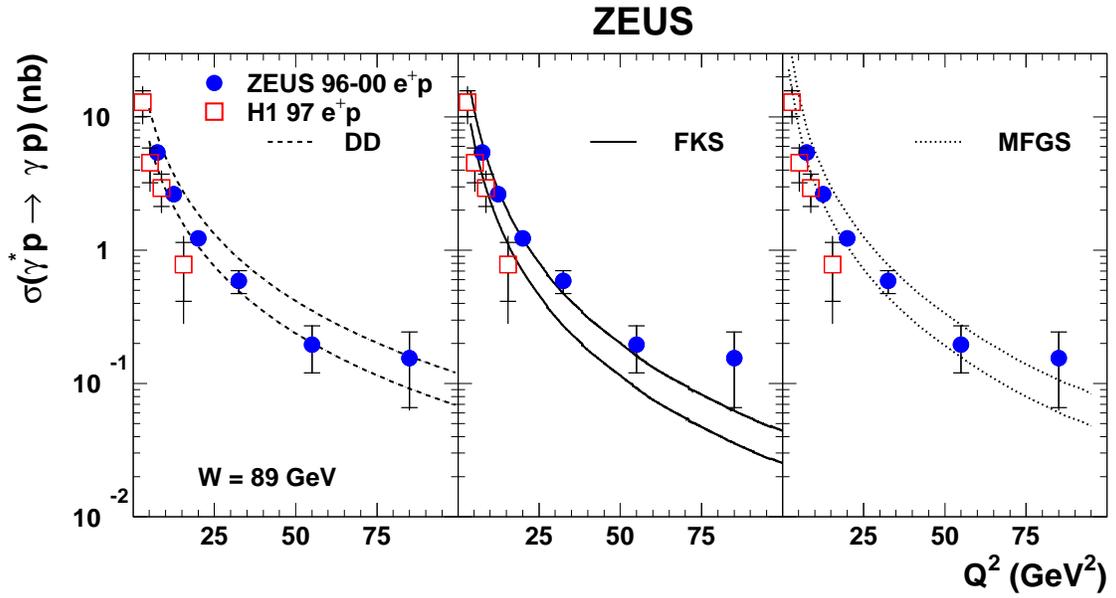,width=0.99\textwidth,bb=25 29 753 444}
\caption{The DVCS cross section, $\sigma(\gamma ^*p\rightarrow \gamma
p)$, as a function of $Q^2$, as also shown in
Fig.~\protect\ref{fig:q2_models}, for the ZEUS (dots) and H1 (squares)
data. The data are compared to the theoretical predictions of the DD,
FKS and MFGS models of colour-dipole interactions.  The curves
correspond to fixed $b$ values, $b=4\;\Gev^{-2}$ (upper) and
$b=7\;\Gev^{-2}$ (lower).
\label{fig:cdm_q2}}
\end{figure}

The DVCS cross section has also been calculated within colour-dipole
models~\cite{zfp:c49:607,zfp:c53:331,np:b415:373,np:b425:471,Favart:2003cu},
which have been successful in describing both the inclusive and the
diffractive DIS cross sections at high
energy~\cite{pr:d59:014017,pr:d60:114023,pr:d60:074012,pl:b425:369,
epj:c16:641,plb:b502:74}. The various dipole models differ in their
formulation of the dipole cross section with the target proton. If
$s$-channel helicity is conserved in DVCS, the virtual photon must be
transversely polarised.  As the wave function of the
transversely-polarised photon can select large dipole sizes, whose
interactions are predominantly soft, DVCS constitutes a good probe of
the transition between perturbative and non-perturbative regimes of
QCD.  The $Q^2$ dependence of $\sigma_{\rm{DVCS}}$ has been compared
to the expectations of three calculations based on colour-dipole
models, by Donnachie and Dosch (DD)~\cite{plb:b502:74}, Forshaw,
Kerley and Shaw (FKS)~\cite{pr:d60:074012,np:a675:80c,epj:c22:655} and
McDermott, Frankfurt, Guzey and Strikman
(MFGS)~\cite{epj:c22:655,epj:c16:641}. The comparisons are shown in
Fig.~\ref{fig:cdm_q2}, where the model expectations are represented by
curves corresponding to a fixed value of $b=4\;\Gev^{-2}$ (upper) and
$b=7\;\Gev^{-2}$ (lower), chosen for illustration.  All three
predictions give a reasonable representation of the data.  The H1
measurements~\cite{pl:b517:47} are also shown, extrapolated to the $W$
value of the ZEUS data using the $W^\delta$ dependence of the cross
section measured in this analysis for the $e^+p$ data. The H1 data lie
systematically below the ZEUS data.

%%%%%%%%%%%%%%%%%%%%%%%%%%%%%%%%%%%%%%%%%%%%%%%%%%%%%%%%%%%%%%%%
\section{Conclusions}
\label{sec-conclusions}
%%%%%%%%%%%%%%%%%%%%%%%%%%%%%%%%%%%%%%%%%%%%%%%%%%%%%%%%%%%%%%%%%

The DVCS cross section $\sigma(\gamma^\ast p \rightarrow \gamma p)$
has been measured at HERA in the kinematic range \mbox{$5<Q^2<100
  \,\Gev^2$} and \mbox{$40<W<140 \,\Gev$}. No significant difference
between the $e^+p$ and $e^-p$ interactions was observed.  The data
have been compared to calculations based on generalised parton
distributions and on the colour-dipole model. Generally, good
agreement with the data is observed.

The $Q^2$ dependence of the DVCS cross section follows approximately a
$Q^{-3}$ behaviour. The precision of the data allows an accurate
determination of the $W$ distribution for the first time.  The cross
section rises steeply with $W$, indicative of a hard underlying
process, where the rise reflects the increase of parton distributions
with decreasing Bjorken $x$.

These measurements demonstrate the potential of DVCS data to constrain
the structure of the proton and quark-gluon dynamics at low $x$.

%%%%%%%%%%%%%%%%%%%%%%%%%%%%%%%%%%%%%%%%%%%%%%%%%%%%%%%%%%%%%%%%
\section*{Acknowledgements}
\label{sec-ack}
%%%%%%%%%%%%%%%%%%%%%%%%%%%%%%%%%%%%%%%%%%%%%%%%%%%%%%%%%%%%%%%%%

We thank the DESY directorate for their strong support and
encouragement. The special efforts of the HERA machine group are
gratefully acknowledged.  We are grateful for the support of the DESY
computing and network services. The design, construction and
installation of the ZEUS detector have been made possible by the
ingenuity and effort of many people who are not listed as authors.  We
are grateful to M.~Diehl, A.~Freund, M.~McDermott and M.~Strikman for
numerous discussions in which they offered invaluable insight into the
DVCS process. We thank M.~McDermott and R.~Sandapen for providing the
model calculations.

We dedicate this paper to the memory of Bill Toothacker whose early
work on the analysis of DVCS helped us to obtain the preliminary
results that were shown at the EPS meeting in Tampere, Finland, July
1999.

%------------------------------------------------------------------------------
%       Bibliography
%------------------------------------------------------------------------------
{
\def\bibname{\Large\bf References}
\def\refname{\Large\bf References}
\pagestyle{plain}
\ifzeusbst
  \bibliographystyle{./BiBTeX/bst/l4z_default}
\fi
\ifzdrftbst
  \bibliographystyle{./BiBTeX/bst/l4z_draft}
\fi
\ifzbstepj
  \bibliographystyle{./BiBTeX/bst/l4z_epj}
\fi
\ifzbstnp
  \bibliographystyle{./BiBTeX/bst/l4z_np}
\fi
\ifzbstpl
  \bibliographystyle{./BiBTeX/bst/l4z_pl}
\fi
{\raggedright
\bibliography{./BiBTeX/user/syn.bib,%
              ./BiBTeX/bib/l4z_articles.bib,%
              ./BiBTeX/bib/l4z_books.bib,%
              ./BiBTeX/bib/l4z_conferences.bib,%
              ./BiBTeX/bib/l4z_dvcs.bib,%
              ./BiBTeX/bib/l4z_h1.bib,%
              ./BiBTeX/bib/l4z_misc.bib,%
              ./BiBTeX/bib/l4z_old.bib,%
              ./BiBTeX/bib/l4z_preprints.bib,%
              ./BiBTeX/bib/l4z_replaced.bib,%
              ./BiBTeX/bib/l4z_temporary.bib,%
              ./BiBTeX/bib/l4z_zeus.bib}}
}
\vfill\eject

%------------------------------------------------------------------------------
%       Tables
%------------------------------------------------------------------------------
%-------------------------------------------------------------------------------
%       An example table
%-------------------------------------------------------------------------------
\begin{table}[p]
\begin{center}
\begin{tabular}{|c|c|c|c|}
\hline
\multicolumn{4}{|c|}{$\sigma ^{\gamma^*p\rightarrow \gamma p} $}\\
\hline
$Q^2$ range ($\rm{GeV^2}$) & $Q^2$ ($\rm{GeV^2}$) &
 $\sigma ^{\gamma^*p\rightarrow \gamma p}$ (nb)&$\sigma ^{\gamma^*p\rightarrow \gamma p}$ (nb) \\
&& $e^+p$&$e^-p$\\ 
\hline
5 - 10 & 7.5& 5.42$\pm$0.33$_{-0.34}^{+0.29}$   &5.63$\pm$0.77$^{+0.30}_{-0.33}$\\
10 - 15 & 12.5 & 2.64$\pm$0.22$_{-0.13}^{+0.11}$& 2.20$\pm$0.52$^{+0.13}_{-0.14}$\\
15 - 25 & 20 & 1.23$\pm$0.14$_{-0.07}^{+0.05}$  & 0.96$\pm$0.31$^{+0.10}_{-0.06}$\\
25 - 40 & 32.5 & 0.59$\pm$0.12$_{-0.04}^{+0.04}$& 0.61$\pm$0.28$^{+0.06}_{-0.05}$\\
40 - 70 & 55 & 0.20$\pm$0.08$_{-0.02}^{+0.03}$&$-$\\
70 - 100 & 85 & 0.16$\pm$0.09$_{-0.03}^{+0.02}$&$-$\\
\hline
\end{tabular}
\caption{Values of the cross sections for the 
  $\gamma ^*p\rightarrow \gamma p$ DVCS process as a function of $Q^2$
  for the $e^+p$ and $e^-p$ data. 
 Values are quoted at the centre of
  each $Q^2$ bin and for the average $W$ value of the whole sample,
  $W=89 \gev$, obtained from GenDVCS. 
 The first uncertainty is
  statistical and the second systematic.  The systematic uncertainty
  due to the luminosity determination is not included.}
  \label{tab:tab1}
\end{center}
\end{table}
%------------------------------------------------------------------------------
\begin{table}[p]
\begin{center}
\begin{tabular}{|c|c|c|c|c|c|}
\hline
\multicolumn{6}{|c|}{$\sigma ^{\gamma^*p\rightarrow \gamma p} $}\\
\hline
$W$ range ($\rm{GeV}$) & $W$ ($\rm{GeV}$) &
 $\sigma ^{\gamma^*p\rightarrow \gamma p} $ (nb)
&$W$ range ($\rm{GeV}$) & $W$ ($\rm{GeV}$) &
$\sigma ^{\gamma^*p\rightarrow \gamma p} $ (nb)\\
$e^+p$ & $e^+p$ &$e^+p$ &$e^-p$ &$e^-p$ &$e^-p$ \\
\hline
40 - 50 & 45 &2.19$\pm$0.24$_{-0.14}^{+0.11}$&&& \\
50 - 60 & 55 &2.96$\pm$0.28$_{-0.18}^{+0.13}$&&& \\
60 - 70 & 65 &3.62$\pm$0.36$_{-0.23}^{+0.18}$&40 - 73&56.7& 2.94$\pm$0.39$^{+0.16}_{-0.13}$\\
70 - 80 & 75 &3.88$\pm$0.42$_{-0.26}^{+0.18}$&&& \\
80 - 90 & 85 &3.59$\pm$0.45$_{-0.25}^{+0.18}$&&& \\
90 - 100 & 95 &3.29$\pm$0.55$_{-0.20}^{+0.21}$&73 - 107&90 &4.06$\pm$0.69$^{+0.35}_{-0.25}$ \\
100 - 110 & 105 &6.24$\pm$0.77$_{-0.49}^{+0.31}$&&& \\
110 - 120 & 115 &4.86$\pm$0.76$_{-0.44}^{+0.39}$&&& \\
120 - 130 & 125 &4.69$\pm$0.82$_{-0.36}^{+0.32}$&107 - 140&123.3 &3.8$\pm$1.1$^{+0.3}_{-0.4}$ \\
130 - 140 & 135 & 5.55$\pm$0.99$_{-0.30}^{+0.91}$&&&\\
\hline
\end{tabular}
\caption{Values of the cross sections for the $\gamma ^*p\rightarrow \gamma p$ DVCS process as a function of $W$ for the $e^+p$ and $e^-p$ data.
  Values are quoted at the centre of each $W$ bin and for the average
  $Q^2$ value of the whole sample, $Q^2=9.6 \gev^2$, obtained from
  GenDVCS.  The first uncertainty is statistical and the second
  systematic.  The systematic uncertainty due to the luminosity
  determination is not included.}
  \label{tab:tab3}
\end{center}
\end{table}
%------------------------------------------------------------------------------
\begin{table}[p]
\begin{center}
\begin{tabular}{|c|c|c|c|c|}
\hline
\multicolumn{5}{|c|}{$\sigma ^{\gamma^*p\rightarrow \gamma p} $}\\
\hline
$W$ range ($\rm{GeV}$) & $W$ ($\rm{GeV}$) &
 $\sigma ^{\gamma^*p\rightarrow \gamma p} $ (nb)
&$\sigma ^{\gamma^*p\rightarrow \gamma p} $ (nb) &
$\sigma ^{\gamma^*p\rightarrow \gamma p} $ (nb)\\
& & $5<Q^2<8\gev^2$ & $8<Q^2<13\gev^2$ & $13<Q^2<30\gev^2$\\
& & $Q^2=6.2\;\Gev^2$&$Q^2=9.9\;\Gev^2$ &$Q^2=18.0\;\Gev^2$\\
\hline
40 - 65 &52.5& 5.63$\pm$0.58$^{+0.40}_{-0.35}$ & 2.52$\pm$0.26$^{+0.09}_{-0.18}$ & 0.99$\pm$0.13$^{+0.05}_{-0.10}$\\
65 - 90 &77.5& 6.57$\pm$0.91$^{+0.47}_{-0.81}$ & 3.12$\pm$0.39$^{+0.21}_{-0.17}$ & 1.34$\pm$0.17$^{+0.05}_{-0.09}$ \\
90 - 115 &102.5& 9.5$\pm$1.5$^{+0.8}_{-1.4}$ & 3.94$\pm$0.61$^{+0.32}_{-0.30}$ & 1.91$\pm$0.30$^{+0.12}_{-0.12}$ \\
115 - 140 &127.5& 7.6$\pm$1.6$^{+1.5}_{-0.6}$ & 5.83$\pm$ 0.89$^{+0.49}_{-0.48}$& 1.64$\pm$0.47$^{+0.13}_{-0.15}$\\
\hline
\end{tabular}
\caption{Values of the cross sections for the $\gamma ^*p\rightarrow \gamma p$ DVCS process as a function of $W$ for the $e^+p$ data in three $Q^2$ ranges.
  Values are quoted at the centre of each $W$ bin and for the average
  $Q^2$ values obtained from GenDVCS.  The first uncertainty is
  statistical and the second systematic.  The systematic uncertainty
  due to the luminosity determination is not included.}
  \label{tab:tab4}
\end{center}
\end{table}

%------------------------------------------------------------------------------
%       Figures
%------------------------------------------------------------------------------
%-------------------------------------------------------------------------------
%       Results
%-------------------------------------------------------------------------------

%------------------------------------------------------------------------

%-----------------------------------------------------------------------------

%\begin{figure}
%\centering
%\epsfig{file=Figures/xs_comb_ele_pos_sep_q2.eps,width=0.99\textwidth}
%\caption{$\sigma(\gamma ^*p\rightarrow \gamma p)$ DVCS cross-section
%  as a function of $Q^2$ for $W=89\;\Gev$,
%  separately for $e^+p$ (dots) and $e^-p$ (triangles) interactions.  
%  The solid line is the result of a fit of the form 
%  $\sigma_{DVCS} \sim(Q^2)^{-n}$ to the positron data.  
%  The error bars denote the
%  statistical uncertainty (inner) and the quadratic sum of the
%  statistical and the systematic uncertainties (outer).
%  The $e^-p$ data points are displaced horizontally for ease of visibility.
%\label{fig:e_p_q2}}
%\end{figure}

%----------------------------------------------------------------------------

%----------------------------------------------------------------------------

%----------------------------------------------------------------------------

%---------------------------------------------------------------------------
%\begin{figure}
%\centering
%\epsfig{file=Figures/xs_pos_mcdermott_q2.eps,width=0.99\textwidth}
%\caption{$\sigma(\gamma ^*p\rightarrow \gamma p)$ 
%as a function of $Q^2$, as shown in Figure~\protect\ref{fig:e_p_q2}, compared
%to the GPD-based theoretical predictions of~\protect\cite{misc:fms:gpd}, 
%where MRSTL(M) stands for the LO (NLO) parametrization of parton distribution 
%functions. The MRSTM expectations are also shown for $Q^2$-dependent $b$
%values described in the text.}
%\label{fig:gpd_q2}
%\end{figure}

%----------------------------------------------------------------------------

%
%       ... that's it
%
\end{document}